%% file: NEXT_SiPMs_JINST_2.9.tex
\pdfoutput=1
\documentclass{JINST}
\input{defs.tex}

\title{Design and characterization of the SiPM tracking system of NEXT-DEMO, a demonstrator prototype of the NEXT-100 experiment}
\author{
V.~\'Alvarez,$^{a}$
M.~Ball,$^{a}$
F.I.G.M.~Borges,$^{b}$
S.~C\'arcel,$^{a}$
J.M.~Carmona,$^{c}$
J.~Castel,$^{c}$
J.M.~Catal\'a,$^{d}$
S.~Cebri\'an,$^{c}$
A.~Cervera,$^{a}$
D.~Chan,$^{e}$
C.A.N.~Conde,$^{b}$
T.~Dafni,$^{c}$
T.H.V.T.~Dias,$^{b}$
J.~D\'iaz,$^{a}$
M.~Egorov,$^{e}$
R.~Esteve,$^{d}$
P.~Evtoukhovitch,$^{f}$
L.M.P.~Fernandes,$^{b}$
P.~Ferrario,$^{a}$
A.L.~Ferreira,$^{g}$
E.~Ferrer-Ribas,$^{h}$
E.D.C.~Freitas,$^{b}$
A.N.C. Garcia,$^{b}$
V.M.~Gehman,$^{e}$
A.~Gil,$^{a}$
I.~Giomataris,$^{h}$
A.~Goldschmidt,$^{e}$
H.~G\'omez,$^{c}$
J.J.~G\'omez-Cadenas,$^{a}$\thanks{Spokesperson (gomez@mail.cern.ch)}~
K.~Gonz\'alez,$^{a}$
D. Gonz\'alez-D\'iaz,$^{c}$
R.M.~Guti\'errez,$^{i}$
J.~Hauptman,$^{j}$
J.A.~Hernando Morata,$^{k}$
D.C.~Herrera,$^{c}$
V.~Herrero,$^{d}$
F.J.~Iguaz,$^{h}$
I.G.~Irastorza,$^{c}$
V.~Kalinnikov,$^{f}$
D.~Kiang,$^{e}$
L.~Labarga,$^{l}$
I.~Liubarsky,$^{a}$
J.A.M.~Lopes,$^{b}$
D.~Lorca,$^{a}$\thanks{Corresponding author}~
M.~Losada,$^{i}$
G.~Luz\'on,$^{c}$
A.~Mar\'i,$^{d}$
J.~Mart\'in-Albo,$^{a}$
A.~Mart\'inez,$^{a}$
T.~Miller,$^{e}$
A.~Moiseenko,$^{f}$
F.~Monrabal,$^{a}$
C.M.B.~Monteiro,$^{b}$
J.M.~Monz\'o,$^{c}$
F.J.~Mora,$^{c}$
L.M. Moutinho,$^{g}$
J.~Mu\~noz Vidal,$^{a}$
H.~Natal da Luz,$^{b}$
G.~Navarro,$^{i}$
M.~Nebot,$^{a}$
D.~Nygren,$^{e}$
C.A.B.~Oliveira,$^{eg}$
R.~Palma,$^{m}$
J.~P\'erez,$^{n}$
J.L.~P\'erez Aparicio,$^{m}$
J.~Renner,$^{e}$
L.~Ripoll,$^{o}$
A.~Rodr\'iguez,$^{c}$
J.~Rodr\'iguez,$^{a}$
F.P.~Santos,$^{b}$
J.M.F.~dos Santos,$^{b}$
L.~Segui,$^{c}$
L.~Serra,$^{a}$
D.~Shuman,$^{e}$
C.~Sofka,$^{p}$
M.~Sorel,$^{a}$
J.F.~Toledo,$^{d}$
A.~Tom\'as,$^{c}$
J.~Torrent,$^{o}$
Z.~Tsamalaidze,$^{f}$
D.~V\'azquez,$^{k}$
E.~Velicheva,$^{f}$
J.F.C.A.~Veloso,$^{g}$
J.A.~Villar,$^{c}$
R.C.~Webb,$^{p}$
T.~Weber,$^{e}$
J.~White$^{p}$
and N.~Yahlali$^{a}$\thanks{Corresponding author}~
\\
\llap{$^{a}$}
Instituto de F\'isica Corpuscular (IFIC), CSIC \& Universitat de Val\`encia\\
Calle Catedr\'atico Jos\'e Beltr\'an, 2, 46980 Paterna, Valencia, Spain\\
\llap{$^{b}$}
Departamento de Fisica, Universidade de Coimbra\\
Rua Larga, 3004-516 Coimbra, Portugal\\
\llap{$^c$}
Laboratorio de F\'isica Nuclear y Astropart\'iculas, Universidad de Zaragoza\\ 
Calle Pedro Cerbuna, 12, 50009 Zaragoza, Spain\\
\llap{$^d$}
Instituto de Instrumentaci\'on para Imagen Molecular (I3M), Universitat Polit\`ecnica de Val\`encia\\ 
Camino de Vera, s/n, Edificio 8B, 46022 Valencia, Spain\\
\llap{$^{e}$}
Lawrence Berkeley National Laboratory (LBNL)\\
1 Cyclotron Road, Berkeley, California 94720, USA\\
\llap{$^{f}$}
Joint Institute for Nuclear Research (JINR)\\
Joliot-Curie 6, 141980 Dubna, Russia\\
\llap{$^{g}$}Institute of Nanostructures, Nanomodelling and Nanofabrication (i3N), Universidade de Aveiro\\
Campus de Santiago, 3810-193 Aveiro, Portugal\\
\llap{$^{h}$}IRFU, Centre d'\'Etudes Nucl\'eaires de Saclay (CEA-Saclay)\\
91191 Gif-sur-Yvette, France\\
\llap{$^{i}$}
Centro de Investigaciones en Ciencias B\'asicas y Aplicadas, Universidad Antonio Nari\~no\\ 
Carretera 3 este No.\ 47A-15, Bogot\'a, Colombia\\
\llap{$^{j}$}
Department of Physics and Astronomy, Iowa State University\\
12 Physics Hall, Ames, Iowa 50011-3160, USA\\
\llap{$^{k}$}
Instituto Gallego de F\'isica de Altas Energ\'ias (IGFAE), Univ.\ de Santiago de Compostela\\
Campus sur, R\'ua Xos\'e Mar\'ia Su\'arez N\'u\~nez, s/n, 15782 Santiago de Compostela, Spain\\
\llap{$^{l}$}
Departamento de F\'isica Te\'orica, Universidad Aut\'onoma de Madrid\\
Campus de Cantoblanco, 28049 Madrid, Spain\\
\llap{$^{m}$}
Dpto.\ de Mec\'anica de Medios Continuos y Teor\'ia de Estructuras, Univ.\ Polit\`ecnica de Val\`encia\\
Camino de Vera, s/n, 46071 Valencia, Spain\\
\llap{$^{n}$}
Instituto de F\'isica Te\'orica (IFT), UAM/CSIC\\
Campus de Cantoblanco, 28049 Madrid, Spain\\
\llap{$^{o}$}
Escola Polit\`ecnica Superior, Universitat de Girona\\
Av.~Montilivi, s/n, 17071 Girona, Spain\\
\llap{$^{p}$}
Department of Physics and Astronomy, Texas A\&M University\\
College Station, Texas 77843-4242, USA\\

 E-mail:  \email{nadia.yahlali@ific.uv.es, david.lorca@ific.uv.es}
}

\abstract{NEXT-100 experiment aims at searching the neutrinoless double-beta decay of the \XE\  isotope using a TPC filled with a 100 kg of high-pressure gaseous xenon, with 90\% isotopic enrichment. The experiment will take place at the Laboratorio Subterráneo de Canfranc (LSC), Spain.
NEXT-100 uses electroluminescence (EL) technology for energy measurement with a resolution better than 1\% FWHM. 
The gaseous xenon in the TPC additionally allows the tracks of the two beta particles to be recorded, which are expected to have a length of up to 30 cm at 10 bar pressure.  
The ability to record the topological signature of the \bbonu\ events provides a powerful background rejection factor for the \bb\ experiment. 

In this paper, we present a novel 3D imaging concept using SiPMs coated with tetraphenyl butadiene (TPB) for the EL read out and its first implementation in NEXT-DEMO, a large-scale prototype of the NEXT-100 experiment.
The design and the first characterization measurements of the NEXT-DEMO SiPM tracking system are presented.
The SiPM response uniformity over the tracking plane drawn from its gain map is shown to be better than 4\%. 
An automated active control system for the stabilization of the SiPMs gain was developed, based on the voltage supply compensation of the gain drifts. The gain is shown to be stabilized within 0.2\% relative variation around its nominal value, provided by Hamamatsu, in a temperature range of 10$^\circ$C.

The noise level from the electronics and the SiPM dark noise is shown to lay typically below the level of 10 photoelectrons (pe) in the ADC. Hence, a detection threshold at 10 pe is set for the acquisition of the tracking signals. The ADC full dynamic range (4096 channels) is shown to be adequate for signal levels of up to 200 pe/$\mu$s, which enables recording most of the tracking signals.  }

\begin{document}
%\linenumbers

%%%%%%%%%%%%%%%%%%%%%%%%%%%%%%% %%%%%%%%%%%%%%%%%%%%%%%%%%%%%%%%%%%%%%%%%%                                      Section 1                                    %%%%%%%%%%%%%%%%%%%
%%%%%%%%%%%%%%%%%%%%%%%%%% %%%%%%%%%%%%%%%%%%%%%%%%%%%%%%%%%%

\section{Introduction}  \label{sec:intro}

NEXT (Neutrino Experiment with a Xenon TPC) will search for the \bbonu\ decay of the {\XE\ }, using a time projection chamber (TPC) filled with 100 kg of high-pressure gaseous xenon, with 90\% isotopic enrichment  \cite{Alvarez:2011,Alvarez:2012-2}. The experiment, so-called NEXT-100, will be held in the Canfranc Underground Laboratory (LSC) \cite{LSC}.
NEXT-100 is designed with the aim of combining an excellent energy resolution with a unique topological signature of the \bb\ events, to achieve high sensitivity to a light Majorana neutrino.  

The main advantage of xenon is a high Q-value of the \XE\ transition (Q$_{\beta\beta}$=2457.83(37) keV \cite{Redshaw:2007}), which minimizes the overlap of the \bbtnu\ and \bbonu\ populations of the \bb\ spectrum. 
In pure xenon, a large amount of primary ionization and primary UV scintillation ($\lambda \sim$ 175 nm) 
result from the interaction of charged particles.
In the TPC, an electric field is used to drift the ionization electrons towards a region where large electroluminescent (EL) signals or secondary scintillation are produced, whose total yield is proportional to the energy of the beta particles. The EL signals can be recorded with negligible electronic noise, which enables the achievement of a near-intrinsic resolution in the energy measurement (< 1\% FWHM at \qbb\ ), i.e. a resolution close to the intrinsic limit determined by the Fano factor that quantifies the fluctuations in the number of ionization electrons in the xenon gas.

The beta particles from the \XE\ $\rightarrow$ $^{136}$Ba decay have tracks of up to 30 cm in gaseous xenon at 10 bar pressure
\cite{Alvarez:2011}. The ability to record these tracks enables a strong suppression (about five orders of magnitude) of the background resulting from the gamma-rays that most severely contaminate the region of interest of the \bbonu\ spectrum \cite{Alvarez:2012-2}. This background originates mainly from the decay of the $^{208}$Tl (2614 keV) and $^{214}$Bi (2447 keV) isotopes that contaminate the detector. The background rate in NEXT-100 is thus expected to be one of the lowest in the new generation of \bbonu\ experiments \cite{GomezCadenas:2011}. 
The design of the high pressure xenon TPC for \bbonu\ searches faces three main challenges :
\begin{itemize}
\item Determination of the total energy of each candidate event with an energy resolution better than 1\% FWHM. This goal has been met in the TPC prototypes presently in operation at LBNL  \cite{Goldschmidt:2011, Alvarez:2012-6} and at IFIC \cite{Ferrario:2012,Alvarez:2012-7}, which performed an energy measurement with a resolution that extrapolates to less than 1\% FWHM at \qbb, assuming 1/$\sqrt{E}$ scaling.
\item Reconstruction of the complete topology of each event in 3D, based on energy-sensitive tracking of the \bb\ decay electrons. The 3D localization requires efficient detection of the primary scintillation light to accurately define the start-of-event time $t_0$.  
\item Selection of materials with high radiopurity which, in conjunction with the background rejection capabilities of the high-pressure xenon gas TPC, provides the desired sensitivity. 
\end{itemize}

\begin{figure}[tbhp!]
\begin{center}
\includegraphics[width=0.75\textwidth]{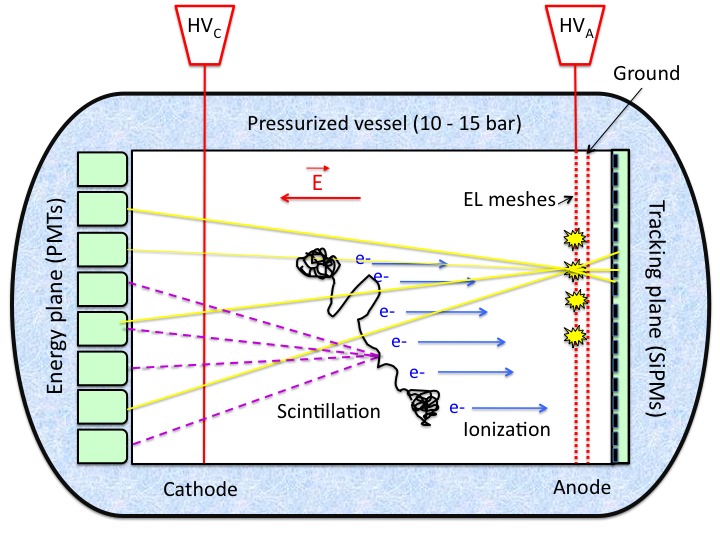} 
\end{center}
\vspace{-0.5cm}
\caption{\small The Separated Optimized Function TPC (SOFT) concept. A tortuous track generates primary scintillation in the gaseous xenon, which is recorded by the array of PMTs located near the TPC cathode. The EL light generated by the primary electrons between the parallel meshes at the anode is recorded in the SiPM plane located right behind them for tracking. It is also recorded in the PMT plane for energy measurements.}
\label{fig:tpc_soft} 
\end{figure}
%%%%%%

The design concept of NEXT TPC is based on specific and separated readout technologies for energy measurement and pattern recognition. 
The energy measurement is provided by an array of low-radioactivity photomultiplier tubes (PMTs) located at the TPC cathode, as illustrated in figure~\ref{fig:tpc_soft}. The tracking measurement is provided by silicon photomultipliers (SiPMs), which provide an optimal spatial resolution at a moderate cost.  
In many applications in medical physics \cite{España:2010} and in particle and nuclear physics, tracking devices based on SiPMs coupled to scintillators have been recently proposed and used \cite{Garutti:2011, Ieki:2009}. These devices are typically segmented in alternating 2D views for a complete 3D track reconstruction. 
The suitability of SiPMs for optical imaging in neutrino and dark matter experiments with noble-gas detectors has been recently demonstrated with the implementation of a 2D imaging concept using SiPMs for EL read out in a large volume liquid argon TPC \cite{Lightfoot:2009, McConkey:2011}. These photosensors offer several advantages for imaging in a large-scale radiopure detector: ruggedness, cost effectiveness and radiopurity at the activity level of some $\mu$Bq/kg for the $^{238}$U and $^{232}$Th radioactive chains \cite{{Heusser:1995},{Alvarez:2012-5}}. On the other hand, due to the insensitivity of the commercial SiPMs to the VUV scintillation of noble gases (peak at 175 nm in xenon and at 129 nm in argon), their use for optical imaging requires shifting the VUV scintillation of the noble-gas to visible light with the wavelength shifter tetraphenyl-butadiene (TPB) \cite{Alvarez:2012-1,Yahlali:2012}. 

In this paper, a novel 3D imaging concept based on SiPMs for EL read out in a high pressure xenon gas TPC is described. This tracking concept is implemented in NEXT-DEMO, the large-scale prototype and demonstrator of NEXT-100 TPC \cite{Ferrario:2012,Alvarez:2012-7}. The full 3D track measurement is accomplished by using a SiPM pixel readout plane (see figure~\ref{fig:tpc_soft}), which detects the EL light produced by the ionization electrons liberated along the tracks, thus providing a 2D imaging of the particle trajectory. The third spatial coordinate is determined by the measurement of the electron drift time, using for the time reference the prompt xenon primary scintillation recorded by the PMT readout plane. 
First 3D track images from cosmic muons, photoelectrons and X-rays using an array of SiPMs for EL read out have been recently obtained in the NEXT prototype operated at LBNL \cite{Alvarez:2012-6}.
The 3D optical imaging concept provides, in principle, a superior pattern recognition performance, which could be a competitive and cost-effective alternative to 3D charge readouts in TPCs for a variety of applications \cite{Delbart:2011}.

In the following sections, we describe the SiPM-based tracking system of NEXT-DEMO prototype. Then we present the measurements performed to characterize the SiPM tracking plane and its readout electronics prior to its in-vessel commissioning.

%%%%%%%%%%%%%%%%%%%%%%%%%%%%%%% %%%%%%%%%%%%%%%%%%%%%%%%%%%%%%%%%%%%%%%%%%                                      Section 2                                    %%%%%%%%%%%%%%%%%%%
%%%%%%%%%%%%%%%%%%%%%%%%%% %%%%%%%%%%%%%%%%%%%%%%%%%%%%%%%%%%

\section{NEXT-DEMO tracking system}
\label{sec:tracking_system}

%%%%%%%%%%%%%%%%%%%%%%%%%%%%%
% Section 2.1 :  Tracking concept
%%%%%%%%%%%%%%%%%%%%%%%%%%%%%

\subsection{Tracking concept} \label{sec:tracking-concept}

\begin{figure}[bthp!]
\begin{center}
\includegraphics[width=0.485\textwidth]{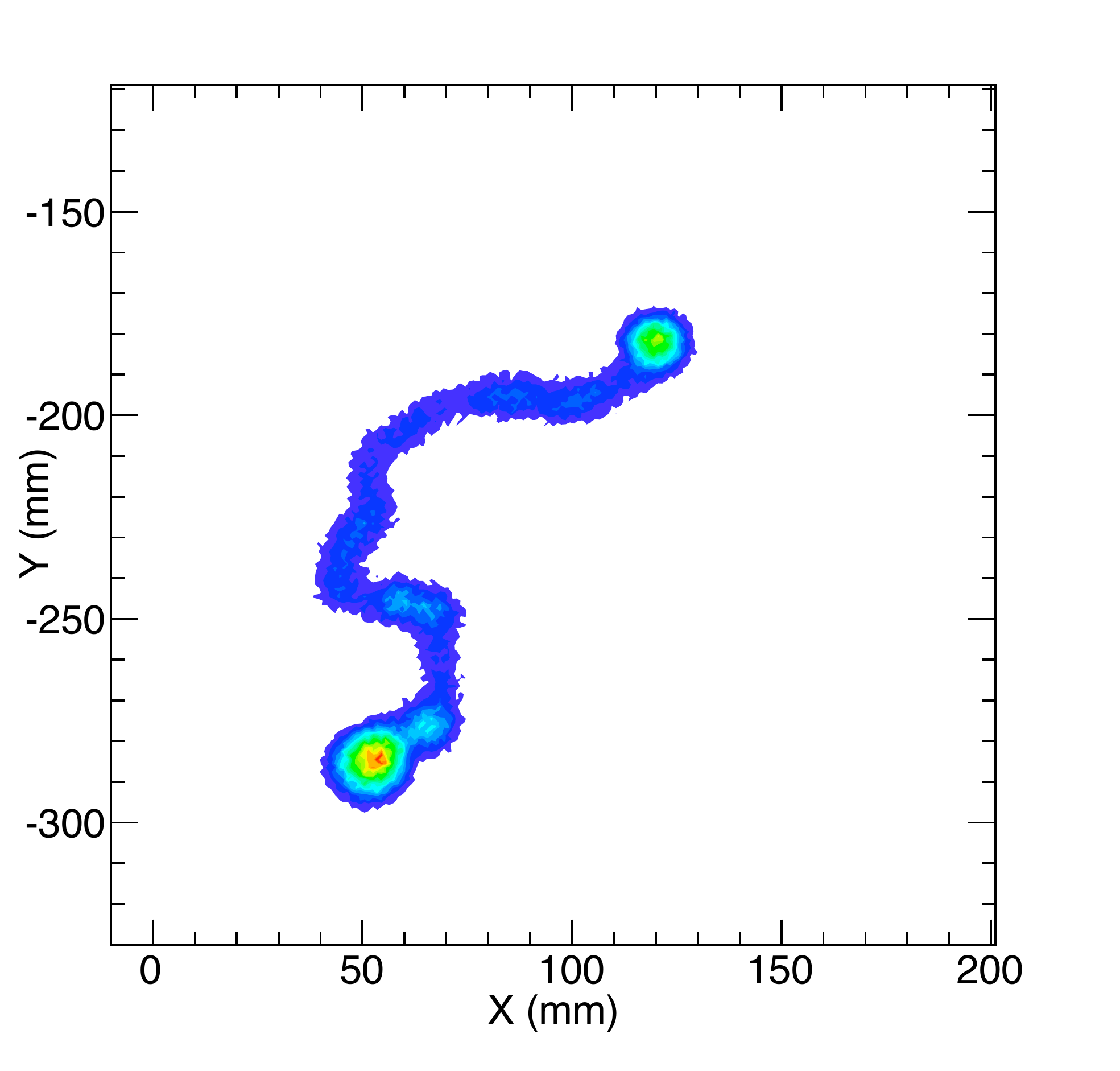}
\includegraphics[width=0.50\textwidth]{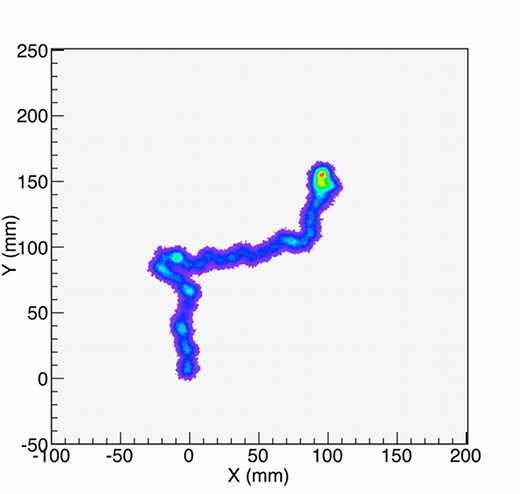}

\caption{Geant4 simulation of the x-y topological signature of a \bb\ event (left), corresponding to two ranging-out electrons \cite{Alvarez:2012-2} and of one-electron event (right), produced by photoelectric effect from the interaction of a $^{214}$Bi gamma-ray (2447 keV) in the xenon gas at 10 bar pressure \cite{Yahlali:2010}.} 
\label{fig:topology}
\end{center}
\end{figure}

A \bbonu\ event will deposit 2458~keV (Q$_{\bb}$ for \XE) in the xenon gas and will produce a track of about 30~cm length at 10 bar pressure.
This track has a distinctive energy deposition pattern in gaseous xenon: a long and tortuous cord due to multiple scattering, ended by two-blobs, corresponding to the ranging-out of the two beta particles 
(see figure~\ref{fig:topology}-(left)). 
The average energy deposition in the track is about 70 keV/cm, except in the blobs at both track ends, where about 200 keV/cm are deposited \cite{Alvarez:2011}. 
The ability to record the topological signature of the \bbonu\ events provides a powerful background rejection factor for the \bb\ experiment \cite{Alvarez:2012-2}.
%%%%%%

%%%%%

The primary scintillation resulting from the particle interaction with the xenon atoms is recorded by the array of PMTs located at the cathode, and defines the start-of-event time t$_{0}$.
The ionization electrons which drift to the TPC anode (at a typical velocity of 1~mm/$\mu$s) generate EL light when crossing the region of intense field (E/p~$\sim$~2.5~kV cm$^{-1}$bar$^{-1}$) between the highly transparent EL meshes.
Thus, setting up a plane with pixel photosensors, located right behind this EL region, enables the measurement of 
the transverse ($x,y$) coordinates of the tracks. The longitudinal coordinate ($z$) is drawn from the time delay {\it t}-{\it t}$_0$ between the time {\it t} at which the EL signals are recorded and the reference time {\it t$_0$}. 

In addition to the position information, the tracking photosensors have to provide a rough measurement of the energy for the reconstruction of the tracks topology. Indeed, the measurement of the energy per unit of length of the tracks is required for the pattern recognition.    
To be able to perform tracking using EL, the detection pixels of a few mm$^{2}$ area should have a limited field of view more-or-less straight towards the parallel EL meshes. 
If the tracking plane is a few mm away from the EL region, the tracking pixels mounted on this plane will produce a signal only while the EL is generated within their field of view. 

The spatial resolution in the tracking plane is naturally limited by the transverse diffusion of the ionization electrons during the drift time. In pure xenon at 10 bar pressure and at a typical drift field of 1~kV/cm, the maximum transverse diffusion in NEXT-100 is of the order of 10 mm, as drawn from \cite{Biagi:1999}. 
This suggests a minimum spacing of 10 mm of the detection pixels in the tracking plane, leading to about 10$^{4}$ channels per m$^2$. 
 
In a \bbonu\ event, the average number of primary electrons released by ionization of the gas along the track is about 300 electrons per mm (assuming an energy deposition of 70 keV/cm and a mean energy to produce an electron-ion pair in xenon gas of $21.7$~eV \cite{Dias:1997}).
For an incident drift velocity of 1~mm/$\mu$s, about 300 ionization electrons from a track directed along the drift direction will enter the EL region every microsecond. Each electron produces EL light for a time interval given by the gap size between the meshes divided by the drift velocity in the EL region. This is several microseconds for 5 mm gap and a reduced EL field of E/p $\sim$ 2.5~kV cm$^{-1}$ bar$^{-1}$.  
For this typical EL field value actually used in NEXT-DEMO prototype, the optical gain is about 1100 \cite{Monteiro:2007}, thus the total number of EL photons produced is on average 3.3$\times10^{5}$ per microsecond.
The tracks less parallel to the drift axis contribute with a higher number of electrons per unit time within the EL gap and the number of EL photons produced can be much higher.  

A detection element of 1~mm$^2$ located at 5 mm distance from the EL region, in the direct view of a track parallel to 
the drift axis, will subtend a mean solid angle fraction of 0.0016. Hence, about 460 photons/$\mu$s will impinge on that detection area, assuming 88\% transparency of the EL meshes.  
If we consider a SiPM of 1~mm$^2$ active area as a detection element, with 50\% maximum photon detection efficiency (PDE), coated with TPB, the effective detection efficiency of the EL photons is about 22\%, assuming 90\% conversion efficiency of the coating and 50\% reemission onto the SiPM surface. This estimation is compatible with the recent measurements of the effective PDE of a TPB-coated SiPM reported in \cite{Yahlali:2012}. 
Hence, a signal level of about 100 photoelectrons (pe) per $\mu$s is expected from SiPMs of 1~mm$^2$ active area and 50\% maximum PDE, instrumenting a tracking plane located 5 mm behind the EL meshes.

In this simple estimate, multiple scattering and diffusion of the drifting electrons are neglected. 
The imaging resolution of the tracking system depends on the distance between the meshes and the tracking plane. The smaller this distance the better is the imaging resolution, provided the illumination level on the SiPMs is below their saturation limit and within the dynamic range of the readout electronics.
In the NEXT-DEMO prototype this distance is set to 10~mm, which is the maximum available space for positioning the tracking plane behind the EL grids in the TPC. 
The information of the 3D coordinates of the tracks provided by the tracking pixels should be completely registered for the analysis of the event topology. This requires that the events are recorded for the entire drift time in the TPC, typically 300~$\mu$s in 
NEXT-DEMO and 1~ms in NEXT-100.

%%%%%%%%%%%%%%%%%%%%%%%%%%%%%
%   Implementation
%%%%%%%%%%%%%%%%%%%%%%%%%%%%%

\subsection{SiPM tracking plane} \label{sec:design}

\begin{figure}[tbhp!]
\begin{center}
\includegraphics[width=0.7\textwidth]{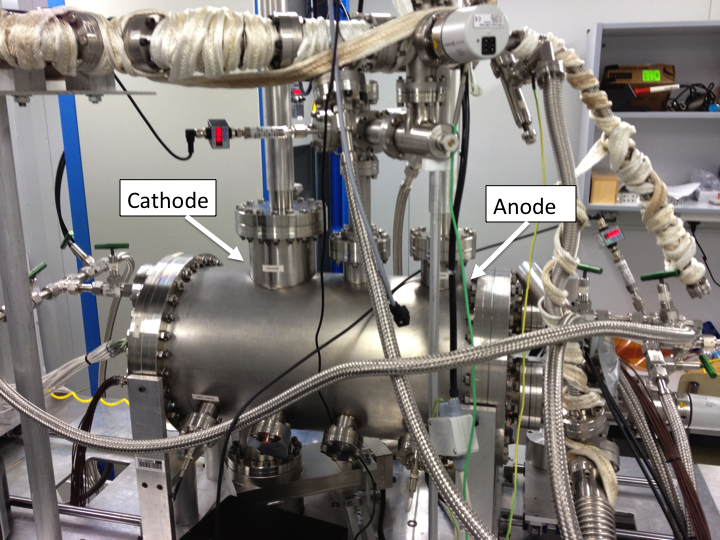} 
\end{center}
\vspace{-0.5cm}
\caption{\small Photograph of the NEXT-DEMO prototype presently in operation at IFIC.}
\label{fig:NEXT-DEMO} 
\end{figure}

The first implementation of the NEXT tracking concept was performed in the NEXT-DEMO prototype (figure~\ref{fig:NEXT-DEMO}). This TPC, containing 1~kg of pure gaseous xenon at 10 bar pressure, has a fiducial volume of 30 cm length and 16 cm inner diameter, which can contain the tracks of about 8 cm length produced by 662 keV X-rays from a $^{137}$Cs radioactive source.

In the first operation phase of NEXT-DEMO, the TPC was operated with 19 pressure-resistant PMTs at the cathode for accurate energy measurements and 19 similar PMTs at the anode for the x-y localization of the events. This first version of the tracking detector, using 1~inch PMTs, was mainly intended to correct the energy spectra for the dependence on the radial position of the events. See \cite{Alvarez:2012-7} for further details. 
%%%
%%%
In the second phase towards the development of the NEXT tracking system, 248 SiPMs of 1~mm$^2$ active area were used for instrumenting the tracking plane. Two MPPC types from Hamamatsu, S10362-11-025P and S10362-11-050P \cite{Hamamatsu}, were considered.  

The outstanding features of the MPPC S10362-11-025P are pixel size of 25~$\mu$m$\times$25~$\mu$m, gain of $2.75\times10^5$  at the nominal voltage of $\approx 71$~V (typical over-voltage of 1.5 V) and at 25$^{\circ}$C, typical photon detection efficiency (PDE) indicated by Hamamatsu of 25\% at 440 nm, and a wide linearity range due to its high number of pixels (1600). The recovery time of the APD pixels of this MPPC indicated by the manufacturer is typically 20 ns. 
The second MPPC considered, S10362-11-050P, has larger pixel size of 50~$\mu$m$\times$50~$\mu$m, higher gain of $7.5\times10^5$ at a similar nominal voltage, and higher PDE of typically 50\% at 440 nm indicated by Hamamatsu, which enable a substantial increase of the photoelectron (pe) resolution in the readout electronics. However, this MPPC has a higher dark noise level and lower linearity range, due to its lower number of pixels (400). The recovery time of the APD pixels in this case is typically 50~ns. The typical PDE values indicated by Hamamatsu are measured using the photocurrent method in which the contribution of optical cross-talk and after-pulses cannot be subtracted \cite{Yahlali:2012, Vacheret:2011}. The true PDE values of the considered MPPCs free of after-pulsing and crosstalk contributions reported in references \cite{Eckert:2010, Vacheret:2011} are about 30\% lower than those indicated by the manufacturer.    

The after-pulsing probability we may expect from the MPPCs considered has been measured and reported in \cite{Eckert:2010}. It is about 2\% for the MPPC S10362-11-025P operated at a typical over-voltage of 1.5 V, and about 7\% for the MPPCs S10362-11-050P operated at a typical over-voltage of 1 V.  The after-pulsing rate may deteriorate the photon-counting resolution of the MPPCs and thus their absolute charge information. However, the determination of the absolute charge truly induced by the EL photons is not required for tracking. The useful tracking information is provided by the relative charge measurement of the SiPMs array. 

The SiPMs of the two types were arranged to cover a circular plane of 16 cm diameter with 1~cm spacing between the photosensors. Two different arrangements and biasing configurations of the SiPMs were developed for the tracking plane, which are presently being tested in NEXT-DEMO with the aim of assessing the development of the NEXT-100 tracking system.
%
%%%%%
\begin{figure}[b]
\begin{center}
\includegraphics[width=0.48\textwidth]{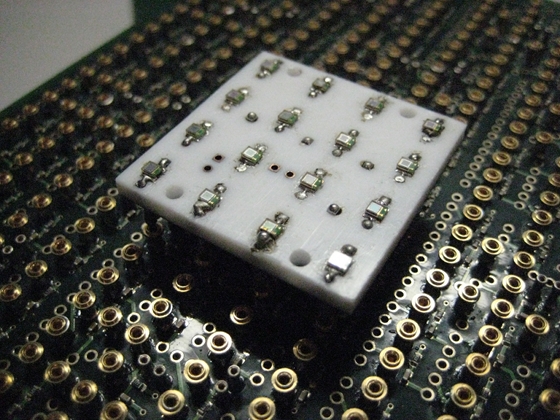}
\includegraphics[height=0.375\textwidth]{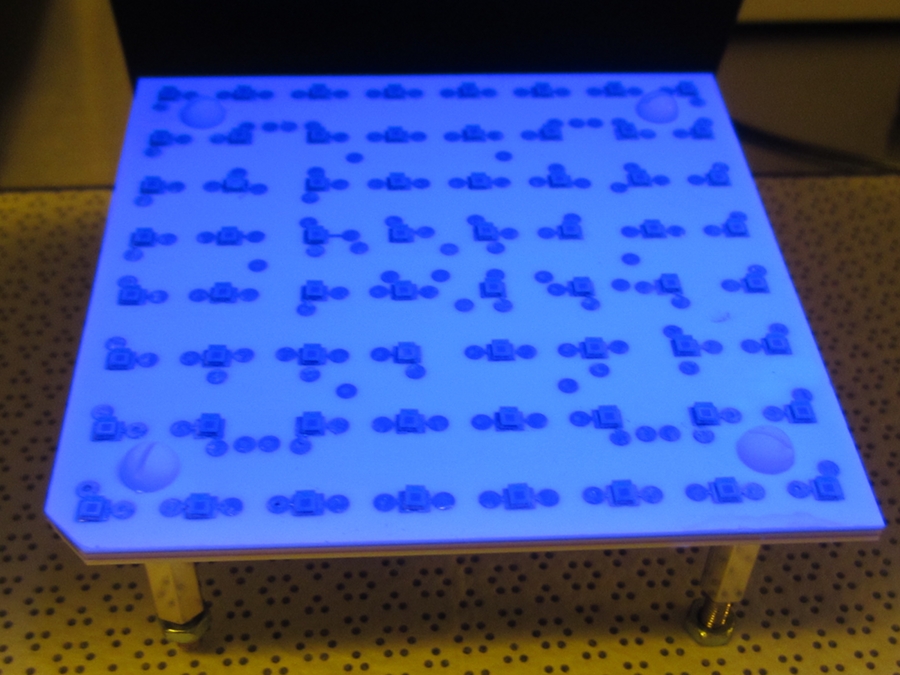}
\end{center}
\caption{(left) Picture of a Cuflon daughter-board with 16 SiPMs, plugged onto the mother board; (right)~picture of a dice-board with 64 SiPMs, coated with TPB and illuminated with 254 nm photons. The converted light from the TPB layer is blue (430 nm). The dice-board contains the electronic circuits and components for biasing the SiPMs.}
\label{fig:DBs}
\end{figure}
%%%%%
%

\begin{figure}[tb]
\begin{center}
\includegraphics[height=0.4\textwidth]{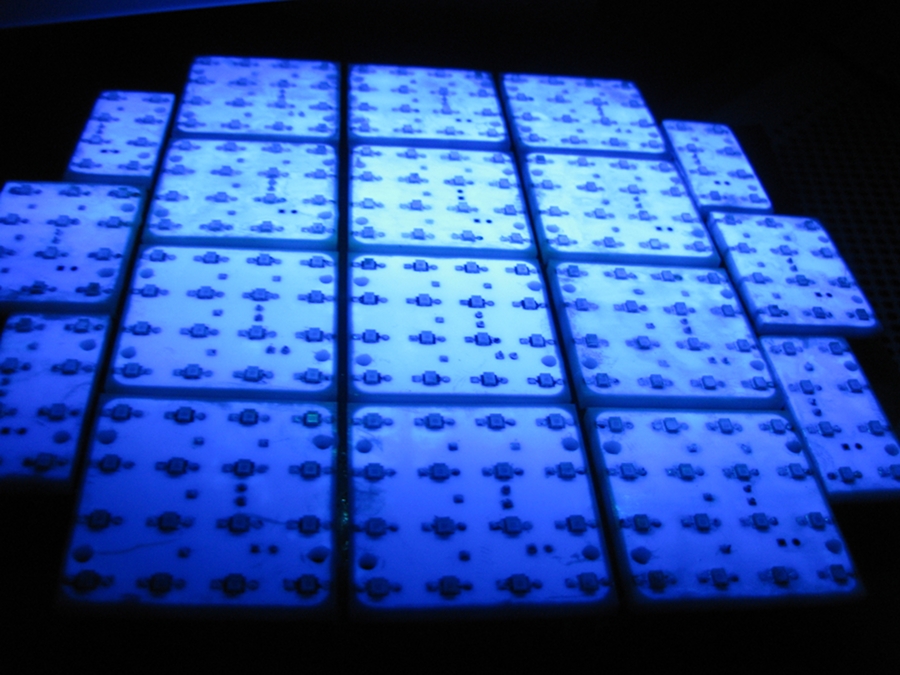}
\end{center}
\caption{Picture of the 18 SiPM daughter-boards of NEXT-DEMO tracking plane, coated with TPB and illuminated with a UV lamp, emitting at 254 nm. Different DB configurations, containing $4\times4$, $2\times4$ or $3\times4$ SiPMs, are arranged to cover the tracking plane.}
\label{fig:tracking-plane}
\end{figure}

The first SiPM arrangement uses the MPPCs of type S10362-11-025P. These were soldered onto 18 daughter-boards (DB) of 38$\times$38 mm$^2$ maximum size (see figure~\ref{fig:DBs}-(left)), made of Cuflon (registered trademark of Polyflon Company  \cite{Polyflon}) which is made of PTFE of 3.18 mm thickness and electroplated with 35 $\mu$m of oxygen-free hard copper. Cuflon has the advantage of high light reflectivity and low degassing. The DBs are plugged onto a mother-board (MB) as shown in figure~\ref{fig:DBs}-(left),
which provides the mechanical support and the electronic circuits for biasing the photosensors. Different DB geometries containing up to 16 SiPMs were built to fit the area of the tracking plane as shown in figure~\ref{fig:tracking-plane}. The DBs were coated with vacuum-evaporated TPB, following the protocol described in \cite{Alvarez:2012-1}. 

The second SiPM arrangement uses the MPPCs of type S10362-11-050P to populate a different version of the Cuflon SiPM board of 78$\times$78 mm$^2$ area, so-called dice-board, shown in figure~\ref{fig:DBs}-(right). The dice-boards comprise 64 SiPMs and the electronic circuits and components for their biasing. These were selected with the requirement of radiopurity, mainly replacing the small ceramic capacitors used for individual SiPMs in the mother-board by large and less radioactive Tantalum capacitors used for groups of 16 SiPMs. Four of these dice-boards units are used to cover the area of the tracking plane in NEXT-DEMO.

The SiPMs in each board were biased using one common operating voltage, taken as the average nominal voltage of all the SiPMs in the board. This biasing option is driven by the necessity of reducing the number of voltage channels and the overall cost of the tracking system due to the high number ($\approx 7000$) of SiPMs in NEXT-100. The common bias introduces, however, a gain dispersion within the SiPM boards, that can be minimized by an adequate selection of the SiPMs as described in section~\ref{sec:gain-calibration}. 
%
%%%%%
%%%%%
%

%%%%%%%%%%%%%%%%%%%%%%%%%%%%%
% Section 2.3 :  Signal processing electronics 
%%%%%%%%%%%%%%%%%%%%%%%%%%%%%

\subsection{Signal processing electronics} \label{sec:electronics}

\begin{figure}[tbh]
\begin{center}
\includegraphics[width=0.8\textwidth]{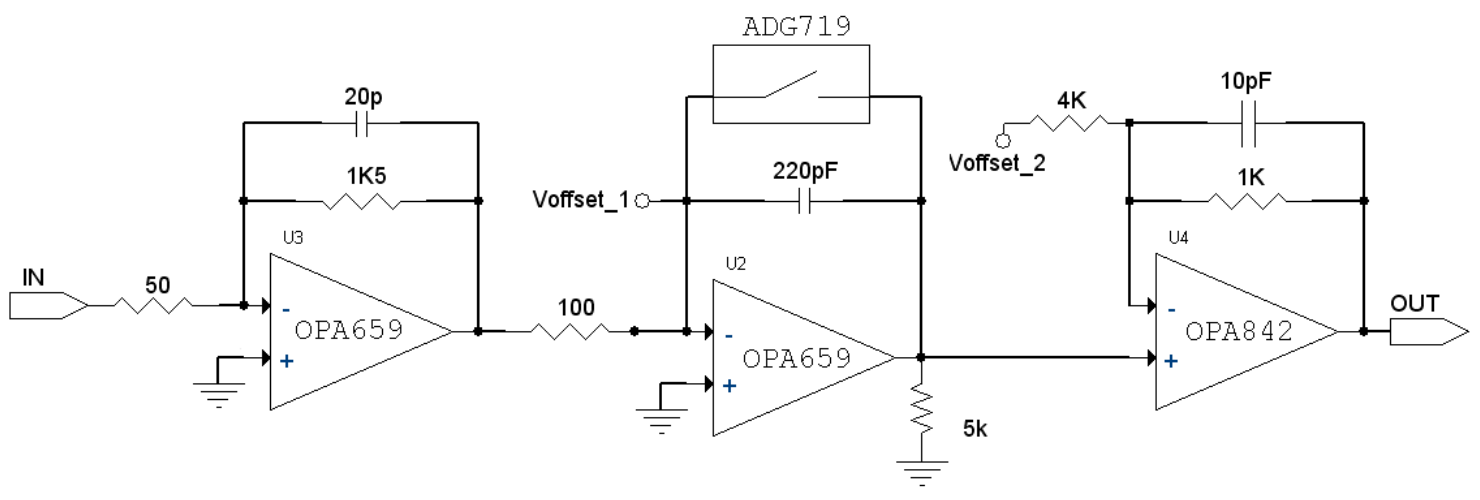}
\includegraphics[width=0.6\textwidth]{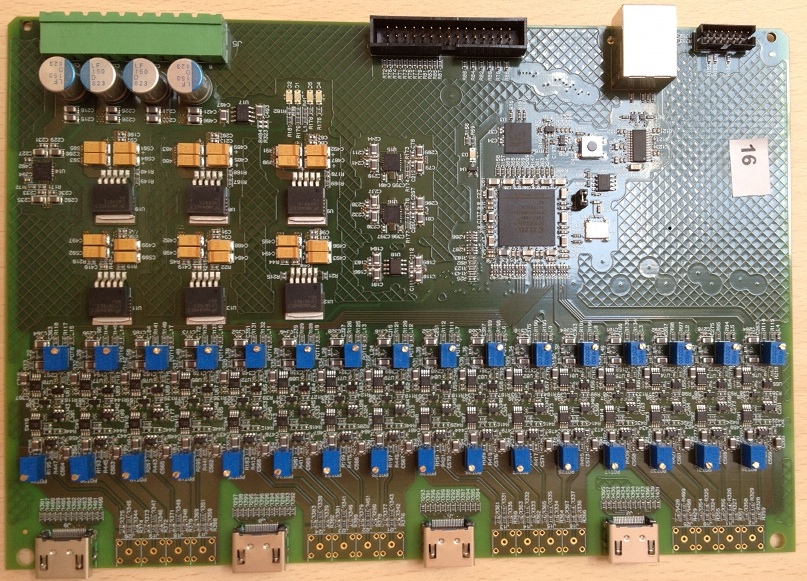}
\end{center}
\caption{(top) Electronic scheme of the analog section of the front-end card; (bottom) picture of the front-end card. }
 \label{fig:FE}
\end{figure}

The typical drift velocity of the electrons along the longitudinal axis ($z$) of the NEXT-DEMO TPC is 1~mm/$\mu$s
\cite{Alvarez:2012-7}. Thus sampling the EL signals at a rate of 1~MHz with an ADC provides a resolution of 1~mm in the $z$ coordinate.
The processing of the SiPM signals is performed by a 16-channel front-end (FE) board (figure \ref{fig:FE}) including 16 analog paths and a digital section. Each analog path consists of three stages. The first stage is a transimpedance amplifier which converts the SiPM current into a voltage signal providing a gain of 1.5 V/mA and baseline adjustment.   
The second stage is a gated integrator with 22 ns RC constant and a nominal integration time of 1~$\mu$s. An offset control at the first stage enables the optimization of the integrator dynamic range.
The third stage is an inverter with a gain of 1.2 required to produce a positive signal at the ADC input. An offset correction is included at this stage since the integrator introduces an output deviation. 
The three electronics stages are manufactured using the OPA659 operational amplifier from Texas Instruments \cite{Texas}. 

The signals obtained in the outputs of the analog paths are digitized at a rate of 1~MHz using 12-bit ADCs (4096 channels). In the digital section, a configurable FPGA (Xilinx Virtex-5 LX50T) is used to read the ADCs, control the switches in the gated integrators (ADG719 in figure \ref{fig:FE}), build a frame with the digitized data and communicate with the upstream readout stage through a standard RJ-45 connector and cable. Careful PCB layout techniques ensure that the digital section introduces very little noise in the analog section. 

%%%%%
%%%%%

The gains in the three stages of the analog section have been set to obtain an output level which can resolve single photoelectrons. For the system to have this resolution the output voltage obtained from a single photoelectron (pe) should be higher than the equivalent output noise of the circuitry which has a typical standard deviation of 2 mV. The gain values at the different stages are set to obtain a voltage level of 17 mV/pe for signals from SiPMs type S10362-11-50P and about 5.6 mV/pe for signals from SiPMs type S10362-11-25P.  The gain values in the analog stages can be further modified in order to adjust the ADCs dynamic range to the level of real tracking signals from the TPC.  

The front-end cards are readout by the Front-End Concentrator (FEC) card, designed within a joint collaboration between CERN-PH-AID and NEXT in the framework of the RD-51 Collaboration \cite{Toledo:2011,Chefdeville:2011}. Up to 16 front-end cards can be connected to the FEC module, resulting in 256 channels, which is enough for the NEXT-DEMO prototype. This readout system can be scaled up for NEXT-100 by simple addition of FEC cards. The data are sent to the data acquisition PC via gigabit Ethernet links. 

The front end electronics in NEXT-DEMO are placed outside the chamber for reasons of space inside the TPC and accessibility of the front end cards for development studies and maintenance. These cards are placed close to the detector in order to minimize signal losses through cables. The upgrade of the front end card presently being developed for the 7000 SiPM channels of  NEXT-100 will include zero suppression and trigger mode readout, allowing a throughput of about 192 Mb/s from the tracking plane for the expected event rate of 10 Hz in the experiment.

%%%%%%%%%%%%%%%%%%%%%%%%%%%%%%% %%%%%%%%%%%%%%%%%%%%%%%%%%%%%%%%%%%%%%%%%%                                      Section 3                                   %%%%%%%%%%%%%%%%%%%
%%%%%%%%%%%%%%%%%%%%%%%%%% %%%%%%%%%%%%%%%%%%%%%%%%%%%%%%%%%%

\section{Characterization of the tracking system} \label{sec:characterization}

The optimal tracking of the \bb\ events using SiPMs requires a uniform and stable response of these photosensors over the tracking area. The behavior of the SiPMs strongly depends on the operating voltage and on temperature. The accurate control of these parameters will ensure the adequate response of the tracking detector in the operation conditions of the TPC.
 The noise level of the SiPMs and of the subsequent electronic circuitry should also be known and should sum up well below the level of the tracking signals. This enables the setting of a detection threshold high enough to efficiently suppress the total noise without affecting the tracking information. 

%%%%%%%%%%%%%%%%%%%%%%%%%%%%%
% Section 3.1 :  Gain map of the tracking plane 
%%%%%%%%%%%%%%%%%%%%%%%%%%%%%

\subsection{Gain map of the tracking plane} \label{sec:gain-calibration}

Due to the Geiger mode operation of the SiPM APD pixels, a relatively small variation of the bias voltage induces a large variation of the SiPM gain (typically $10^5-10^6$) which modifies substantially the counting capability of the photosensor. 
Therefore, in a tracking detector instrumented with SiPMs the gain as a function of bias voltage of each individual photosensor should be well known.
%%%%%%%%%%%%%%%%%%%%%%%%%%%%%%%%%%%%%%%%

In the tracking detector of NEXT-DEMO the 16 SiPMs selected for the population of each of the 18 DBs were chosen to have very close nominal voltages corresponding to a gain value of  2.75$\times10^5$ at 25$^\circ$C. The SiPMs populating each DB were supplied with a common bias voltage taken as the average nominal voltage of the SiPMs on the board. In order to assess the response and uniformity of the photosensors over the tracking plane the gain of the individual SiPMs at their common bias voltage in the corresponding DB was measured. The dispersion of the gain within the 18 DBs, due to the chosen biasing solution, was drawn to evaluate the uniformity level obtained over the tracking area.

\begin{figure}[tb]
\begin{center}
\includegraphics[width=0.50\textwidth]{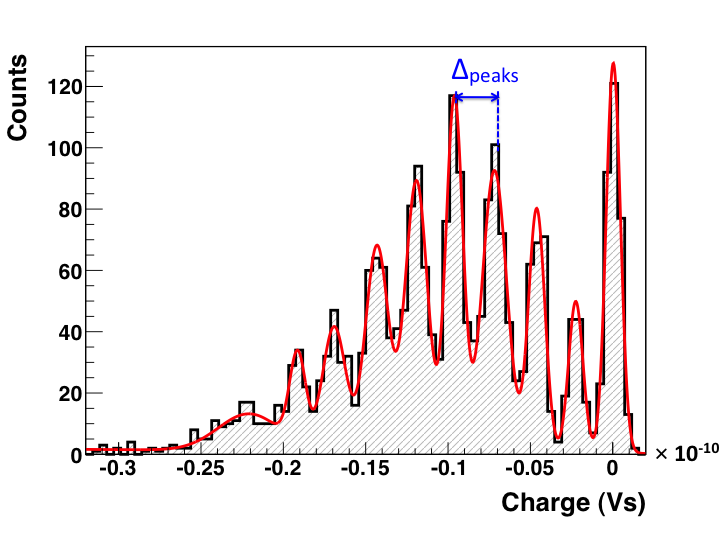}
\includegraphics[width=0.49\textwidth]{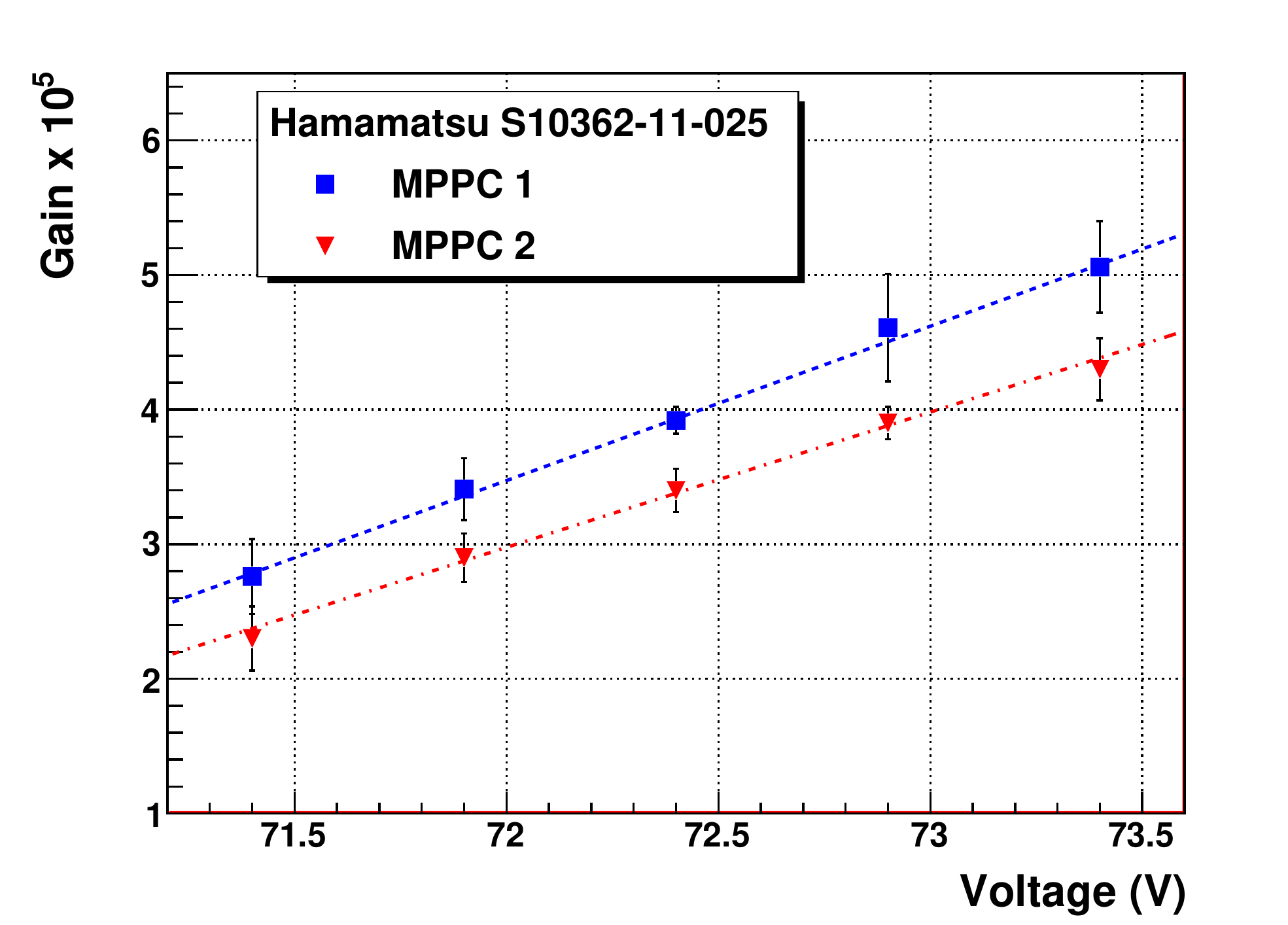}
\end{center}
\caption{(left) Low light spectrum of a SiPM recorded with an oscilloscope, showing the gaussian fits to the photon peaks; (right) gain of two SiPMs (type S10362-11-025P from Hamamatsu) as a function of bias voltage.}
\label{fig:gain_analysis}
\end{figure}

A dedicated setup was used for the gain measurement. The DBs placed in a dark box, were
biased using an electrometer (Keithley 6517B) as a stable voltage source and were illuminated at low intensity with a LED emitting at 400~nm. The LED was triggered with a gate generator (Agilent 33250A) and operated in pulsed mode at 1~kHz and 30~ns pulse width.  The single photon spectra (SPS) of the SiPMs were recorded using an oscilloscope and analyzed to determine their gain at the common bias voltage of the corresponding DB. The measurements were made at room temperature. 
A typical SiPM SPS is shown in figure~\ref{fig:gain_analysis}-(left).  The SiPM gain was obtained from the average charge value between consecutive peaks in the spectrum, which determines the average number of electron-hole pairs produced in the photosensor by a single photon. The gain is given by

\begin{equation}
G=\frac{\Delta_{peaks}}{R \cdot e} 
\end{equation}
%%%%%
%
\noindent
where $\Delta_{peaks}$ is the average distance between consecutive peaks (in Vs), {\it e} is the elementary electron charge and R the input impedance of the oscilloscope. Figure~\ref{fig:gain_analysis}-(right) shows gain curves versus bias voltage of two SiPMs in a range in the vicinities of the nominal voltage recommended by Hamamatsu. As seen, there is a linear dependence of the SiPM gain on the bias voltage in this range.

\begin{figure}[tbhp!]
\begin{center}
\includegraphics[width=0.5\textwidth]{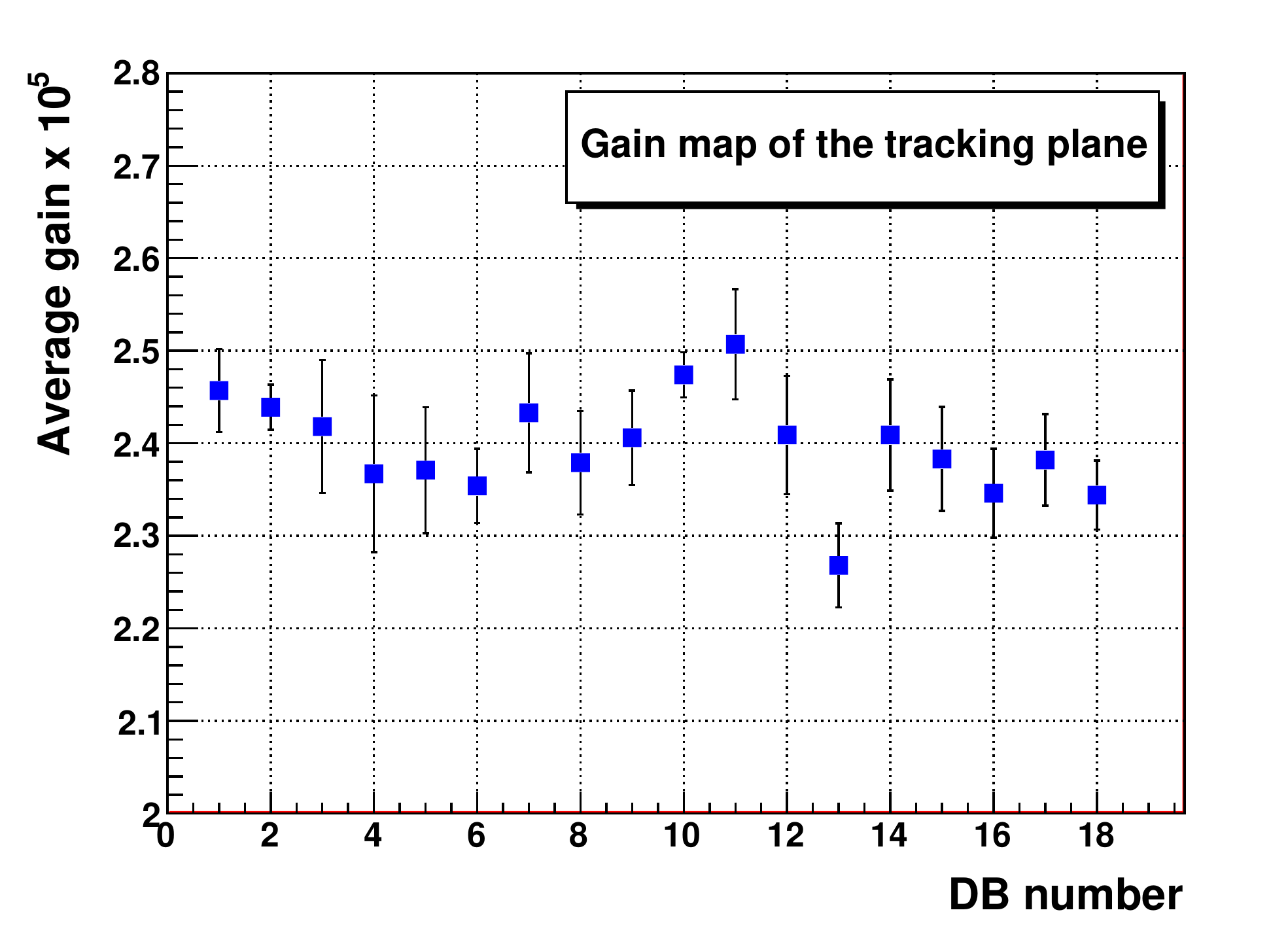}
\includegraphics[width=0.48\textwidth]{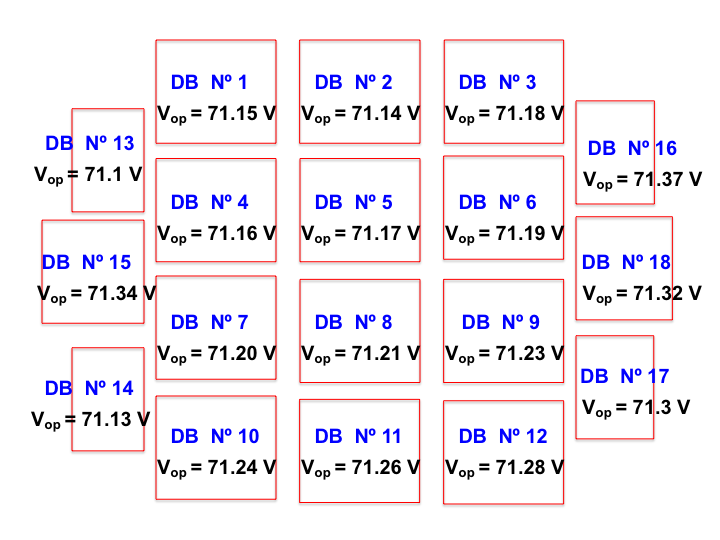}
\caption{(left) Distribution of the SiPM average gain for each DB. The average gain for all the DBs is $2.46\times10^{5}$ with 1.83\% relative standard deviation; (right) configuration of the 18 DBs in the tracking plane facing the electroluminescence grids.}
\label{fig:gain_map}
\end{center}
\end{figure}

The distribution of the average gain in the 18 DBs (248 SiPMs) at their average nominal bias voltage is shown in figure~\ref{fig:gain_map}. The average gain in the set of 18 DBs varies between $2.27\times10^{5}$ (DB\#13) and $2.50\times10^{5}$ (DB\#11), which is lower than the gain specified by Hamamatsu due to the effect of temperature, not corrected in these measurements. 
Indeed, the gain measurements were performed in a black box in which the ambient temperature recorded reached values between 26$^\circ$C and 28$^\circ$C because of the simultaneous operation of various electronic devices nearby. The decrease of the SiPM gain with temperature is typically 2-3\%/$^\circ$C as shown in figure~\ref{fig:gain_control}-(left). This explains the lower gain values measured for the SiPMs of the 18 DBs (average $2.46\times10^5$) with respect to the gain specified by Hamamatsu ($2.75\times10^5$ at 
25$^\circ$C). The gain spread within the DBs varies between 1.0\% (DB\#10) and 3.6\% (DB\#4). This is considered a good enough uniformity level for tracking purposes.

%%%%%
%%%%%

\subsection{Active control of the gain}  \label{sec:gain-control} 

The SiPM gain is very sensitive to temperature variations. For a given value of the reverse bias, the gain is shown to decrease by a factor of 2 for a temperature increase of 10$^\circ$C \cite{Marrocchesi:2009}. 
The thermal agitation in the SiPM dissipates the energy of the charge carriers, which inhibits their collection and decreases the gain at a fixed reverse voltage.
This temperature dependence of the gain can seriously impair the photon counting capability of the photosensors and have a negative impact on the charge measurements in the NEXT tracking detector.

For a stable operation of the SiPMs in NEXT, it is necessary to correct for the gain drifts induced by temperature changes during the detector operation. As an alternative to cooling the photosensors and maintaining them at a very stable temperature, an active control of the gain was implemented using a linear control loop on the operating voltage.  The programmable power supply system developed for the implementation of this bias voltage control is described in \cite{Gil:2012}. 
The temperature at the SiPMs location was measured and used as an input to calculate the appropriate value of the bias voltage for a given nominal gain. This approach requires a precise knowledge of the dependence of the SiPM gain on the bias voltage and on the temperature in the range from 20$^{\circ}$C to 30$^{\circ}$C expected in the laboratory during the detector operation. 

\begin{figure}[tbhp!]
\begin{center}
\includegraphics[width=0.49\textwidth]{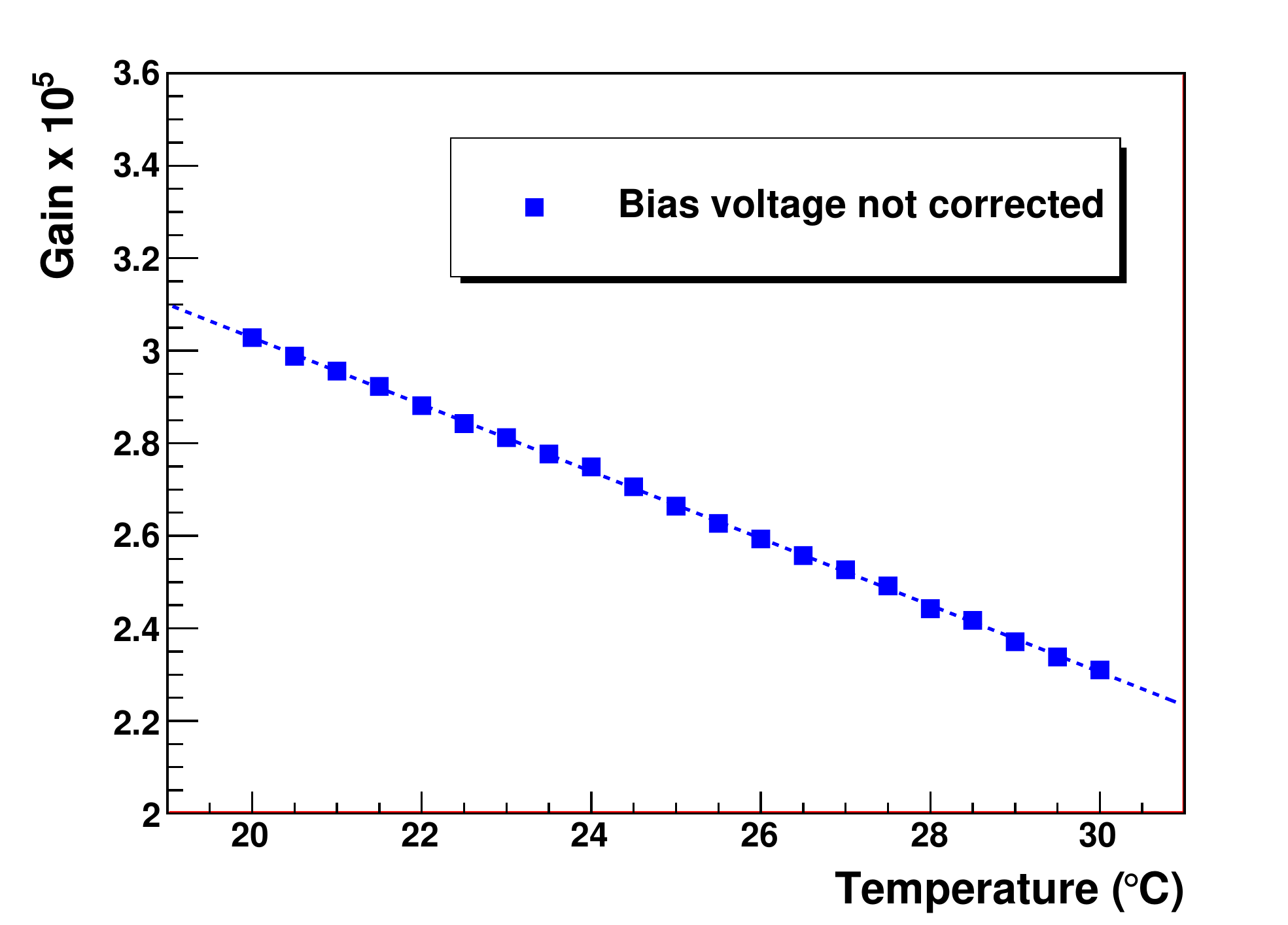}
\includegraphics[width=0.49\textwidth]{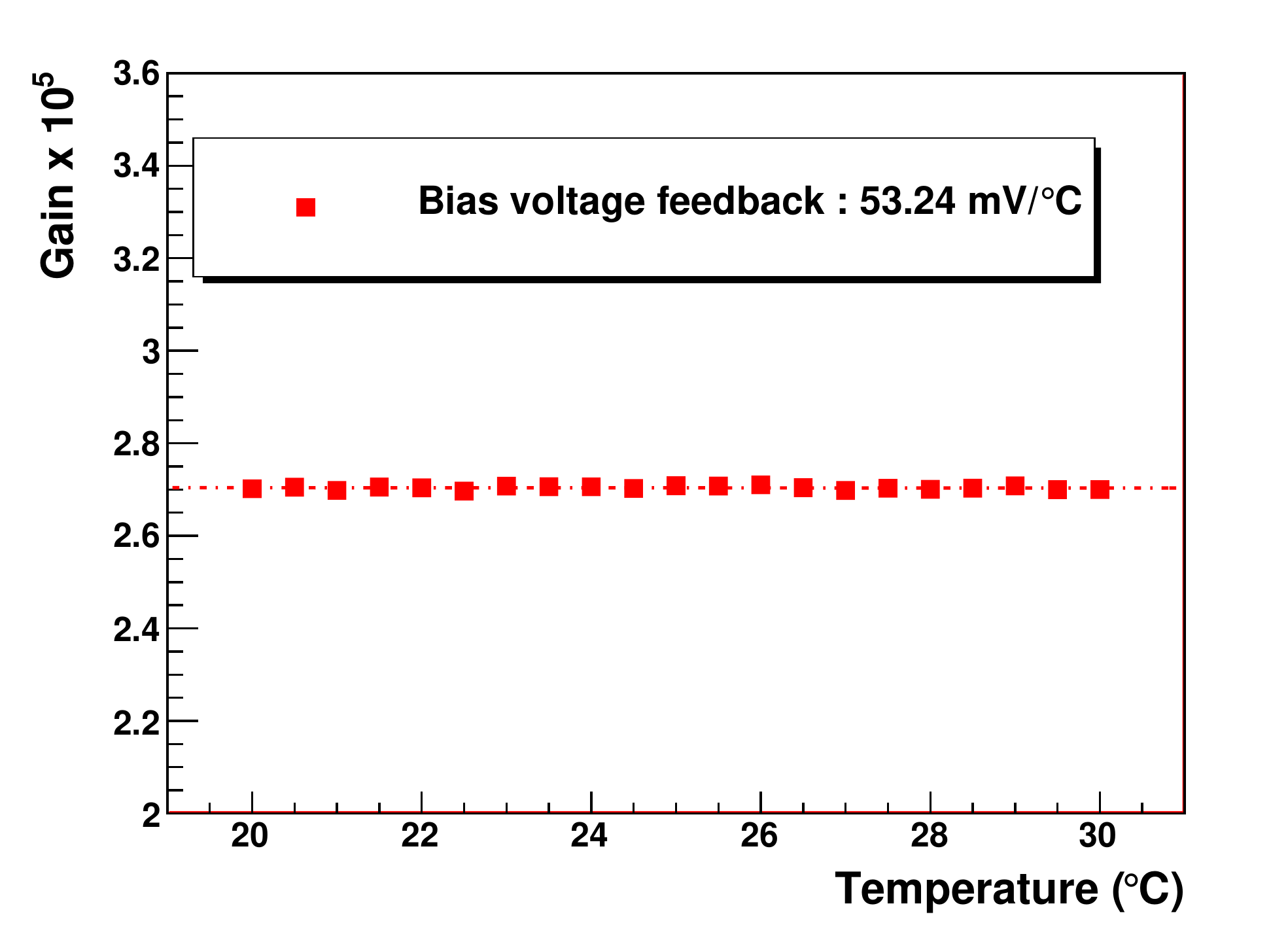}
\caption{
Gain of a SiPM (Hamamatsu S10362-11-25P) as a function of temperature without (left) and with (right) bias voltage correction of 53.24 mV/$^{\circ}$C for the SiPM. The nominal operating voltage provided by Hamamatsu for the tested SiPM is 71.15 V. }
\label{fig:gain_control}
\end{center}
\end{figure}

A dedicated experimental setup was used for the temperature dependence measurements. 
It consisted of a Peltier cell mounted on a copper plate in thermal contact with the SiPM tested. The adjustable bias voltage of the Peltier cell set the temperature in the SiPM, which was measured using a digital thermometer DS18B20 from Maxim Integrated Products Inc.~\cite{Maxim}. This has a  programmable resolution of 9-bit to 12-bit and an accuracy of $\pm 0.5^{\circ}$C over the range $-10^{\circ}$C to $+85^{\circ}$C.
The gain of the SiPM biased at the nominal operating voltage provided by Hamamatsu was measured at different temperatures using their single photon response determined in the experimental setup described in section~\ref{sec:gain-calibration}.
The gain measured decreases linearly with the temperature as shown in figure~\ref{fig:gain_control}-(left). 
The voltage correction that can be applied to the SiPM bias to stabilize the gain is drawn from the known linear dependence of the gain on the voltage and on the temperature. 
The result of this voltage control applied to the SiPM bias    
is shown in figure~\ref{fig:gain_control}-(right), where the stabilized gain is depicted as a function of temperature in the range 
20$^{\circ}$C to 30$^{\circ}$C.
As it can be seen, the SiPM gain is stabilized at its nominal value ($2.7\times10^5$ for the Hamamatsu S10362-11-25P) with a relative standard deviation of 0.14\% in a temperature interval of 10$^{\circ}$C. 
%
%%%%%
%%%%%
%   

In the NEXT-DEMO prototype, the temperature changes in the SiPM plane are expected to be slow and not exceeding a few $^{\circ}$C during the detector operation, due to the continuous gas flow in the TPC and the stable ambient temperature in the laboratory.  Therefore, the stabilization of the SiPM gain, as shown in 
figure~\ref{fig:gain_control}-(right), can be considered accurate enough in the operating conditions of the SiPM tracking detector.
The active control of the gain should however be implemented on the common bias of the SiPM boards containing up to 64 SiPMs. 

The temperature measurement in the tracking detector is provided by digital thermometers DS18B20 with a resolution of 12 bits soldered on each of the SiPM boards composing the tracking detector.
The thermometers are read out in parallel at a frequency of about 1~Hz. The bias correction applied to the SiPM boards to stabilize the SiPMs gain is determined as the average bias correction of the SiPMs composing the board. These are determined precisely from the gain curves as a function of voltage and temperature of the individual SiPMs.
The gain measurements are performed using an automated test system recently developed for the acquisition of up to 64 SiPM single photon spectra. This system consists of the laboratory equipments described above and in section~\ref{sec:gain-calibration} controlled by a software based on LabVIEW, and a custom electronic board with relays for the selection of the individual SiPMs to be tested from a SiPM board. For more details see reference \cite{Rodriguez:2012}. 
For the in-vessel measurements of the SiPM gain, a LED emitting at 240 nm from Roithner LaserTechnic \cite{Roithner} placed at the TPC cathode is used in order to illuminate the tracking plane and enables the gain measurements when required during experimental data taking.  

%%%%%%%%%%%%%%%%%%%%%%%%%%%%%%%%%%%
% Section 3.4 :  Detection threshold and dynamic range 
%%%%%%%%%%%%%%%%%%%%%%%%%%%%%%%%%%%

%\subsection{Detection threshold and dynamic range} \label{sec:dynamic-range}
\subsection{Detection threshold and dynamic range} \label{sec:dynamic-range}

SiPMs are known to be very noisy devices at the single photoelectron level, due to the high rate (about 1~MHz per mm$^2$ active area) of their thermally generated charge carriers at room temperature. This dark noise gives single photoelectron equivalent signals in the SiPMs.  The determination of the pe level of the dark noise from a SiPM of type S1032-11-50P was measured to define a detection threshold for the NEXT-DEMO tracking system.

\begin{figure}[tbhp!]
\begin{center}
\includegraphics[width=0.6\textwidth]{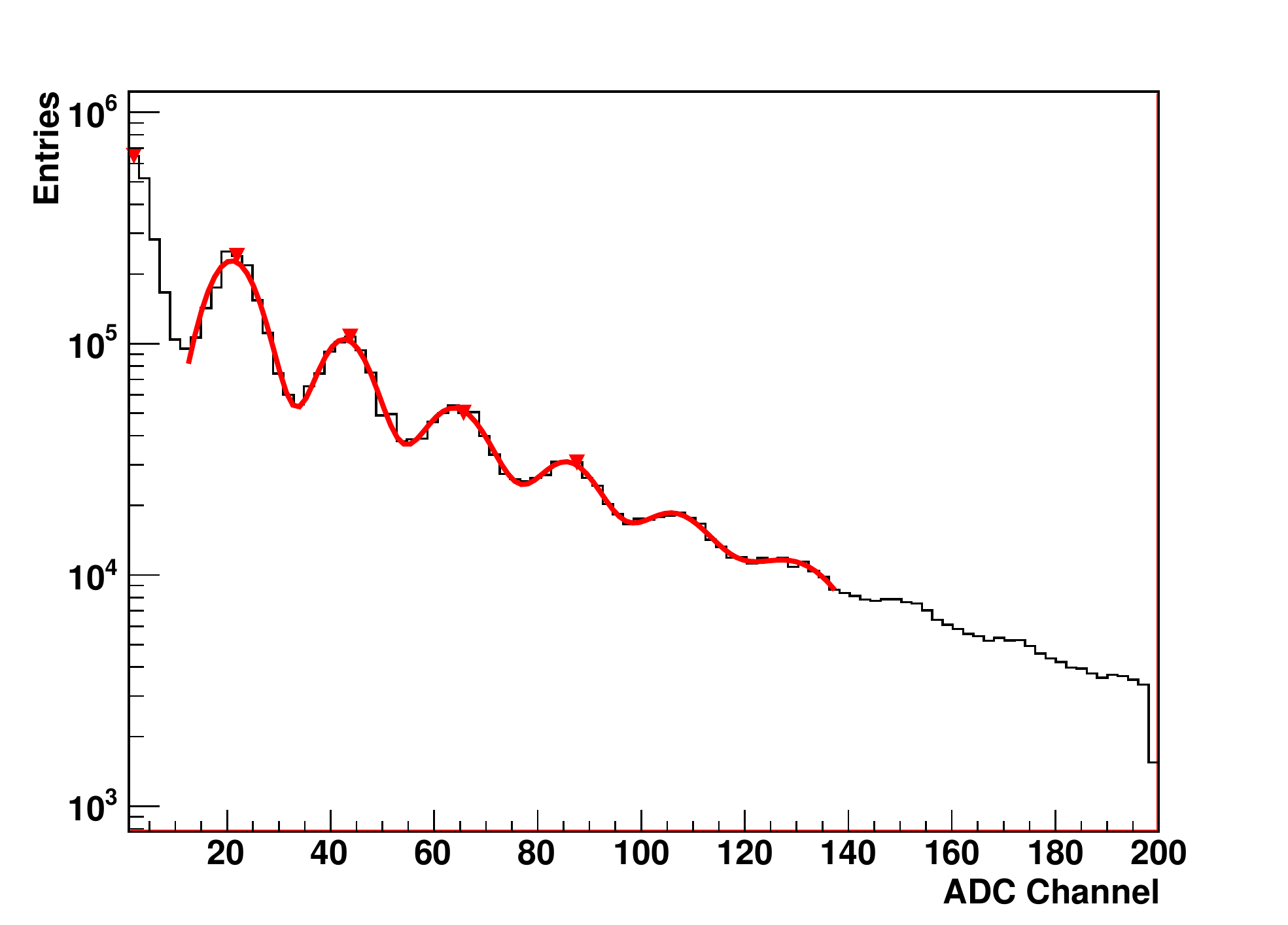} 
\end{center}
\vspace{-0.5cm}
\caption{\small Dark noise spectrum of a SiPM type S10362-11-050P recorded in a sample time of 1~$\mu$s. The gaussian fits to the pe peaks are shown.}
\label{fig:SPS_dark} 
\end{figure}

The SiPM mounted on a dice-board was biased at its nominal operating voltage for dark noise measurements inside a black box. The SiPM signals were read out with the NEXT-DEMO processing electronics described in section~\ref{sec:electronics}. 
The typical charge spectrum from the SiPM dark noise recorded in ADC channels during a sample time of 1~$\mu$s is shown in figure~\ref{fig:SPS_dark}. As can be seen, the single electron charge peaks are well resolved in the interval below 140 ADC channels. At 200 ADC channels, the statistics of the charge peaks from the dark noise decreases by two orders of magnitude and it is negligible at higher ADC channels where the tracking signals are expected to lay. A detection threshold at 200 ADC channels is thus adequate for the read out of EL signals with the SiPMs of type S1032-11-50P. The noise level from the SiPMs of smaller pixel size (type S10362-11-025P) is an order of magnitude smaller. The same detection threshold can thus be set for the tracking plane using these photosensors in NEXT-DEMO.

\begin{figure}[tb]
\begin{center}
\includegraphics[width=0.6\textwidth]{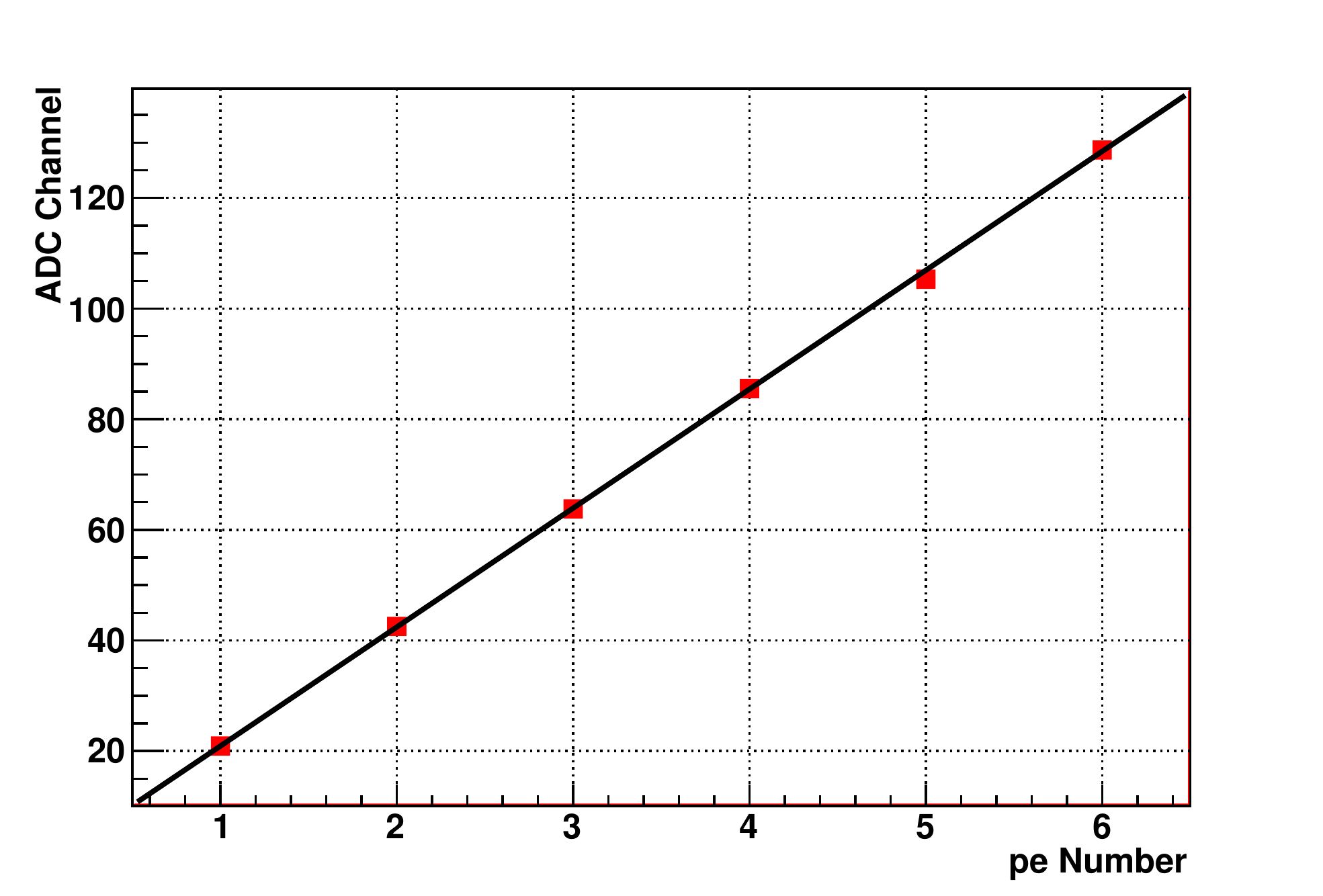}
\end{center}
\caption{Centroid positions of the charge peaks from the SiPM dark noise as a function of the peak number. The slope of the linear fit is $21.50 \pm 0.01$~channels/pe. The errors in the centroid positions drawn from the gaussian fits in the dark noise spectrum 
are small and not visible in the plot.}
\label{fig:ADC_calibration}
\end{figure}

In figure~\ref{fig:SPS_dark} the charge peaks from the dark noise are fitted by gaussians whose centroid positions in ADC channels scale linearly with the number of peaks as shown in figure~\ref{fig:ADC_calibration}.  At the 7th and higher charge peaks the statistics is low, which makes the determination of their centroid position less accurate.  
The ADC integral linearity specified by the manufacturer and checked in the laboratory is of 0.02\% (this is a deviation of one ADC code over the full dynamic range of 4096 codes). This allows the extrapolation of the linear ADC conversion of about 21 channels/pe from below 140 channels to the full ADC dynamic range. 
%%%%%
%
Hence, the detection threshold in NEXT-DEMO is set at 10 pe, and the maximum signal level from the SiPMs that can be measured in the full ADC dynamic range is about 200 pe/$\mu$s. This is adequate for recording most of the tracking signals from the SiPMs in the operating conditions considered in section~\ref{sec:tracking-concept}. 
%
%%%%%
%%%%%
%

The precise determination of the linearity range of the TPB-coated SiPMs selected for the tracking plane, is presently in progress. 
At high illumination levels in the tracking plane, the non-linearity that may occur in the response of the SiPMs is not expected to compromise significantly the tracking information. Indeed, the precise determination of the particle energy deposited per time slices of $1\mu$s is provided by the PMTs in the energy plane, which allow the measurements provided by the SiPMs to be corrected. 

%%%%%%%%%%%%%%%%%%%%%%%%%%%%%%% %%%%%%%%%%%%%%%%%%%%%%%%%%%%%%%%%%%%%%%%%%                                      Section 4                                    %%%%%%%%%%%%%%%%%%%
%%%%%%%%%%%%%%%%%%%%%%%%%% %%%%%%%%%%%%%%%%%%%%%%%%%%%%%%%%%%

\section{Summary and outlook}

The ability to record the event tracks and topology in the NEXT detector is a key feature for the background rejection and the identification of the \bb\ events. 
The present paper describes the design concept of the NEXT tracking system for 3D imaging of the particle tracks in the gaseous xenon TPC, and its implementation in the TPC prototype NEXT-DEMO. This concept based on SiPMs as tracking sensors is thought to provide a competitive alternative to 3D charge readouts in TPCs for a variety of applications.  
 
The first characterization measurements of the NEXT-DEMO tracking plane and its readout electronics are presented.
It is shown in particular that the response of the SiPMs is uniform over the tracking area with a gain dispersion of less 
than 4\%. The gain of the SiPMs is stabilized against temperature changes using an automated bias voltage compensation system, which ensures less than 0.2\% gain variation in a temperature range of 10$^{\circ}$C.
The ADCs are shown to resolve single photoelectrons from the dark noise signals, which can be used for calibration. 
A detection threshold at the level of 10 photoelectrons was shown to reduce significantly the SiPM dark noise contribution to the tracking signals. The full dynamic range of the ADCs is shown adequate for signal levels of up to 200~pe/$\mu$s from the SiPMs. The SiPM tracking plane, developed for 3D optical imaging in NEXT-DEMO, is currently being commissioned.   
 %%%%%%%%%%%%%%%%%%%%%%%
 
 \acknowledgments
We acknowledge the Spanish MICINN for the Consolider Ingenio grants under contracts CSD2008-00037, CSD2007-00042 and for the research grants under contract FPA2008-03456 and FPA2009-13697-C04-01 part of which come from FEDER funds.\\
The Portuguese teams acknowledges support from FCT and FEDER through program COMPETE, projects PTDC/FIS/103860/2008 and PTDC/FIS/112272/2009.\\
J. Renner acknowledges the support of the U.S. Department of Energy Stewardship Science Graduate Fellowship, grant number DE-FC52-08NA28752. 

%%%%%%%%%%%%%%%%%%%%%%%

\end{document}

%% file: defs.tex
\newcommand{\bbonu}{\ensuremath{\beta\beta0\nu}}
\newcommand{\bbtnu}{\ensuremath{\beta\beta2\nu}}
\newcommand{\qbb}{\ensuremath{Q_{\beta\beta}}}

\newcommand{\bb}{\ensuremath{\beta\beta}}

\newcommand{\XE}{\ensuremath{{}^{136}\rm Xe}}